\documentclass[letter,traditabstract]{aa} % for the letters
\usepackage{graphicx}
\usepackage{txfonts}
\begin{document}
   \title{{\it Herschel}\thanks{{\it Herschel} is an ESA space observatory with science instruments provided by European-led Principal
Investigator consortia and with important participation from NASA.}-PACS spectroscopy of IR-bright galaxies at high redshift}

%   \subtitle{I. First results from SHINING}

   \author{E. Sturm\inst{\ref{inst1}}
          \and
          A. Verma\inst{\ref{inst2}}
          \and
          J. Graci\'a-Carpio\inst{\ref{inst1}}
          \and
          S. Hailey-Dunsheath\inst{\ref{inst1}}
          \and
          A. Contursi\inst{\ref{inst1}}
          \and
          J. Fischer\inst{\ref{inst3}}$^,$\thanks{Visiting scientist at Max-Planck-Institute for Extraterrestrial Physics (MPE), Garching, Germany}
          \and
          E. Gonz\'alez-Alfonso\inst{\ref{inst4}}
          \and
          A. Poglitsch\inst{\ref{inst1}}
          \and
          A. Sternberg\inst{\ref{inst5}}
          \and
          R. Genzel\inst{\ref{inst1}}
          \and
          D. Lutz\inst{\ref{inst1}}
          \and
          L. Tacconi\inst{\ref{inst1}}
          \and
          N. Christopher\inst{\ref{inst2}}
          \and
          J. de Jong\inst{\ref{inst1}}
          }

   \institute{Max-Planck-Institute for Extraterrestrial Physics (MPE),
              Gie\ss enbachstra\ss e 1, D-85748 Garching, Germany\\
              \email{sturm@mpe.mpg.de}\label{inst1}
         \and
             Oxford University, Dept. of Astrophysics, Oxford OX1 3RH, U.K.\label{inst2}
         \and
             Naval Research Laboratory, Remote Sensing Division, 4555 Overlook Ave SW, Washington, DC 20375, USA\label{inst3}
         \and
             Universidad de Alcal\'a de Henares, E-28871 Alcal\'a de Henares, Madrid, Spain\label{inst4}
         \and
             Tel Aviv University, Sackler School of Physics \& Astronomy, Ramat Aviv 69978, Israel\label{inst5}
             }

   \date{Received 30 March 2010; accepted 28 April 2010}

\begin{abstract}
{We present {\it Herschel}-PACS observations of rest-frame mid-infrared and far-infrared spectral line emissions from two lensed, ultra-luminous
infrared galaxies at high redshift: \object{MIPS J142824.0+352619} (MIPS J1428), a starburst-dominated system at z = 1.3, and \object{IRAS
F10214+4724} (F10214), a source at z = 2.3 hosting both star-formation and a luminous AGN. We have detected [O\,I]63$\mu$m and [O\,III]52$\mu$m in
MIPS J1428, and tentatively [O\,III]52$\mu$m in F10214. Together with the recent ZEUS-CSO [C\,II]158$\mu$m detection in MIPS J1428 we can for the
first time combine [O\,I], [C\,II] and far-IR (FIR) continuum measurements for photo-dissociation (PDR) modeling of an ultra-luminous
(L$_{IR}\ge$10$^{12}$L$_\odot$) star forming galaxy at the peak epoch of cosmic star formation. We find that MIPS J1428, contrary to average local
ULIRGs, does not show a deficit in [O\,I] relative to FIR. The combination of far-UV flux $G_0$ and gas density $n$ (derived from the PDR models),
as well as the star formation efficiency (derived from CO and FIR) is similar to normal or starburst galaxies, despite the high infrared luminosity
of this system. In contrast, F10214 has stringent upper limits on [O\,IV] and [S\,III], and an [O\,III]/FIR ratio at least an order of magnitude
lower than local starbursts or AGN, similar to local ULIRGs.
%These are first steps for our goal to interpret the strengths of FIR emission lines in the
%context of photo-ionization and photo-dissociation models, and to compare with local analogs, searching for galaxy evolution effects in the ISM.
}
\end{abstract}


   \maketitle
%
%________________________________________________________________

\section{Introduction}

{\it ISO} and {\it Spitzer} observations have shown that the fraction of cosmic star formation in infrared luminous and ultra-luminous galaxies (
LIRGs/ULIRGs: L$_{IR}$=10$^{11}$L$_\odot$ to a few times 10$^{12}$L$_\odot$) increases from a few percent at z$\sim$0 to more than 50\% at z$>$1
(Genzel \& Cesarsky 2000, Soifer et al. 2008). {\it Herschel} offers the first opportunity to study the population of high-redshift infrared bright
galaxies at wavelengths where they emit most strongly, both through photometry and high-resolution spectroscopy.
%Due to their faintness, these objects require very long exposure times in spectroscopy.
In an attempt to probe for evolution of the ISM in the infrared galaxy population and to explore the limits of the PACS spectrometer we are
investing several tens of hours for extremely deep high-resolution FIR spectroscopy of a (small) sample of galaxies with
L$_{IR}>$10$^{12}$L$_\odot$ at redshifts $\ge$ 1. These data will be complemented by and compared to the results from our ongoing {\it Herschel}
spectroscopy of nearby ULIRGs, metal-rich and metal-poor galaxies, and starburst and AGN template objects. We will thus explore a wide range of
parameter space and redshift. In this letter we present the first results on two high redshift objects, demonstrating the capabilities of the PACS
spectrometer in long integrations and high-z spectroscopic observations.
%__________________________________________________________________

\section{Targets}
\label{sect:targets}

We present spectra of two high-redshift galaxies (MIPS J142824.0+352619 and IRAS F10214+4724) that were observed in our initial Science
Demonstration Phase observations from two PACS GT Key Projects: SHINING (PI E. Sturm) and the "Dusty Young Universe" (PI K. Meisenheimer). Both
galaxies are gravitationally lensed bringing them potentially within the sensitivity of the PACS spectrometer in long integrations.

IRAS F10214+4724 (hereafter F10214) is a ULIRG at redshift z=2.2855. It is known to harbor a coeval AGN, with a starbursting host galaxy.
Differential magnification of these components ($\mu$(AGN)/$\mu$(host)=3, Eisenhardt et al. 1996) potentially complicates the interpretation of any
emission detected. Applying a bolometric magnification factor of 12 (Ao et al. 2008) for the host galaxy yields an intrinsic luminosity of
L$_{FIR}\sim$7 x 10$^{12}$ L$_\odot$. Teplitz et al. (2006) presented a mid-infrared ultra-deep spectrum with {\it Spitzer} IRS at low spectral
resolution. Although classified as a Type II AGN, the mid-IR spectrum shows strong silicate emission around 10$\mu$m similar to Type I QSOs, and
very weak (if any) PAH dust features traditionally used as tracers for star formation. Nevertheless, taking differential extinction into account,
the AGN might contribute less than 50\% to the bolometric luminosity of this object.

MIPS J142824.0+352619 (MIPS J1428 hereafter) was discovered in {\it Spitzer}-MIPS images of the NOAO Deep Wide Field Survey Bootes field.
Subsequent {\it Spitzer}-IRS low resolution spectra (Desai et al. 2006), sub-mm (Borys et al. 2006, Iono et al. 2006, Hailey-Dunsheath et al. 2010)
and VLA observations (Higdon et al. 2005) revealed that it is an extremely luminous, starburst-dominated galaxy at z=1.325. This galaxy shows no
signs of an unobscured or obscured AGN. It is magnified by a foreground galaxy, with a relatively small magnification factor of $\mu\le$8 (Iono et
al. 2006).

%__________________________________________________ One column table
\begin{table*}
\caption{Observation details and measured line fluxes}             % title of Table
\label{table:1}      % is used to refer this table in the text
\centering                          % used for centering table
%\begin{tabular}{l l l l c@{} c c c c}        % centered columns (4 columns)
\begin{tabular}{l l l l c c c c c }        % centered columns (4 columns)
\hline\hline                 % inserts double horizontal lines
Source      & line       & OD & OBSID & n$_{rep}\times$n$_{cyc}^{\mathrm{a}}$ & aor duration & flux$^{\mathrm{b}}$       & continuum flux  & corr.$^{\mathrm{c}}$ \\    % table heading
            &            &    &       &                                    & [sec] & [10$^{-18}$W/m$^2$] & density [mJy]  &                       \\
\hline                        % inserts single horizontal line
 MIPS J1428 & [O III] 52 & 205 & 1342187779  &  2$\times$7 &  5348 & 3.7 (0.8)    & 117 (35)     & 1/0.62 \\
 MIPS J1428 & [O I] 63   & 205 & 1342187779  &  1$\times$7 &  2730 & 7.8 (1.9)   & 168 (50)     & 1/0.54  \\
 MIPS J1428 & [O III] 88 & 205 & 1342187779  &  7$\times$7 &  18004 &--$^{\mathrm{d}}$& --$^{\mathrm{d}}$ & --  \\
 F10214     & [O IV] 26  & 179 & 1342186812  & 10$\times$4 &  24827 &$\le$6$^{\mathrm{e}}$ & 330 (100) & 1/0.69  \\
 F10214     & [S III] 33 & 185 & 1342187021  &  6$\times$4 &  5864 &$\le$3$^{\mathrm{e}}$ & 378 (115)& 1/0.65  \\
 F10214     & [O III] 52 & 179 & 1342186812  &  \multicolumn{2}{c}{simult. with [O IV]} & 0.9 (0.3)& 445 (130) & 1/0.47     \\
\hline                                   %inserts single line
\end{tabular}

\begin{list}{}{}
\item[$^{\mathrm{a}}$] {\footnotesize Number of line/range repetitions per nod cycle $\times$ number of nod cycles; $^{\mathrm{b}}$
line and continuum fit uncertainty in brackets; an additional calibration uncertainty of 30$\%$ applies; $^{\mathrm{c}}$ Correction factor, applied
to line fluxes and continuum flux densities, to account for PSF losses of the central spatial pixel (see text); $^{\mathrm{d}}$ The signal in this
PACS wavelength range (i.e. $\ge$190$\mu$m) is contaminated by a 2nd order leak, making estimates of upper limits and RMS unreliable;
$^{\mathrm{e}}$ assuming a Gaussian profile with 3-$\sigma$ peak height and FWHM=300km/s.}
\end{list}
\end{table*}

%______________________________________________ Gamma_1 (lg rho, lg e)
   \begin{figure}
   \centering
   \includegraphics[width=7cm]{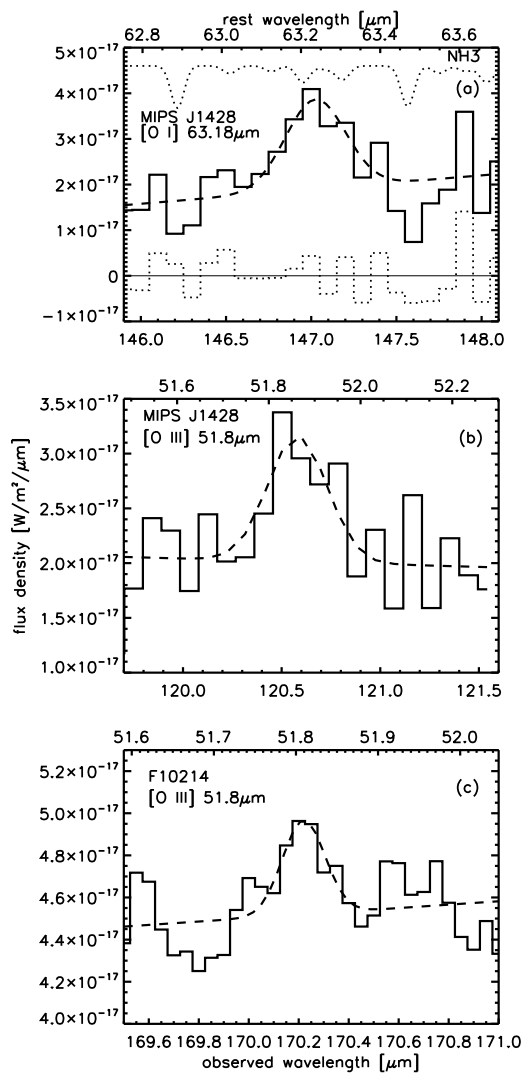}
   \caption{
   Observed PACS spectra. {\it Top:} [O I]63.2$\mu$m in MIPS J1428.
   Overplotted is a Gaussian profile on top of a linear continuum fit. Also shown (upper dotted line) is a NH$_3$ model (arbitrarily scaled), and
   the residual (lower dotted line) after subtracting the line+continuum+NH$_3$ fit from the data.
   {\it Middle:} [O III]51.8$\mu$m in MIPS J1428. Overplotted is a gauss profile on top of a linear continuum fit.
   {\it Bottom:} [O III]51.8$\mu$m in F10214. Overplotted is a gauss+linear continuum fit.}
              \label{Fig1}%
    \end{figure}

\section{Observations and Data Reduction}

The observations were taken with the PACS spectrometer (Poglitsch et al. 2010) on board the {\it Herschel Space Observatory} (Pilbratt et al. 2010)
in high resolution range spectroscopy mode. Most of the data reduction was done using the standard PACS reduction and calibration pipeline (ipipe)
included in HIPE 2.0 1340 \footnote{HIPE is a joint development by the Herschel Science Ground Segment Consortium, consisting of ESA, the NASA
Herschel Science Center, and the HIFI, PACS and SPIRE consortia.}, together with some additional steps to correct for imperfect offset compensation
by the chop/nod observing technique, which can be important at these faint signal levels. The continuum in each of the 16 spectral pixels was
scaled to the median value to correct for residual flat field effects.
%the fact that each spectral pixel sees a slightly different sky position.
Finally the two nod positions were combined to completely remove the sky (telescope) background. Given that our targets are point sources to PACS
we have measured line fluxes from the spectrum in the central spatial pixel of the 5x5 pixel FOV of the PACS spectrometer, applying beam size
correction factors and an additional in-flight correction of the absolute response (1/1.1 in the red and 1/1.3 in the blue PACS bands) as currently
recommended for PACS (Poglitsch et al. 2010). The applied beam size correction factors are listed in Table \ref{table:1}.

\section{Results and Discussion}

We have detected [O\,I]63$\mu$m and [O\,III]52$\mu$m in MIPS J1428, and tentatively [O\,III]52$\mu$m in F10214. The spectra are shown in Figure
\ref{Fig1} (after re-binning to approximately the spectral resolution element, which is appropriate for resolved lines). These are the first
detections of these lines in galaxies at such redshifts, i.e. at the epoch of the peak of cosmic star formation. Flux values, upper limits and
continuum flux densities are given in Table \ref{table:1}. Flux uncertainties from the calibration and from uncertainties in the continuum
definition and line shapes are of the same order, and we estimate a total flux uncertainty of 40-50\%.
%The values in Table \ref{table:1} are not corrected for magnification.

\begin{figure}
   \centering
   \includegraphics[width=7cm]{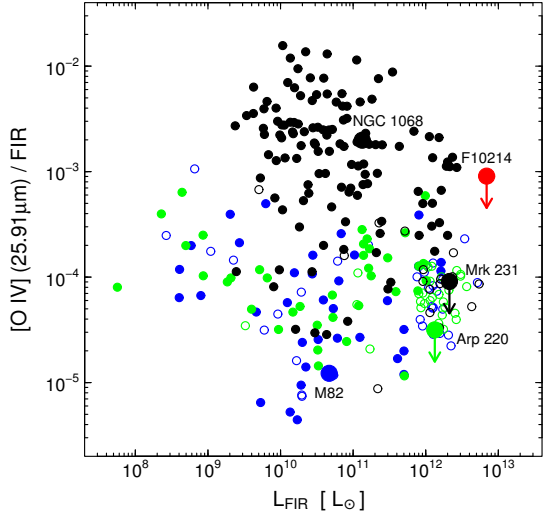}
   \caption{The [O\,IV]/FIR limit in F10214 compared to template objects compiled from the literature (Graci\'a-Carpio et al.,  in prep.).
   Black symbols: AGN and ULIRGs known to harbour an AGN;
   blue: HII galaxies; green: LINERs. Open symbols and arrows are upper line flux limits.
   }
              \label{Fig2}%
    \end{figure}

{\bf F10214:} we did not detect the targeted lines of [S\,III]33$\mu$m and [O\,IV]26$\mu$m, but [O III]52$\mu$m (which is observed in the 1st order
PACS band in parallel with the 2nd order [O IV] data) is serendipitously detected at a $\sim$3$\sigma$-level. We resolve the tentative line and
measure a FWHM of 300$\pm$100km/s (corrected for the instrumental profile, which has a FWHM$\sim$220km/s at 170$\mu$m), similar to the average CO
line width (246$\pm$10 km/s, Ao et al. 2008). Figure \ref{Fig2} shows the ratio of the limit on [O\,IV] to
FIR\footnote{FIR(42-122$\mu$m)=1.26$\times$10$^{-14}$(2.58$\times f60 + f100$) [Wm$^{-2}$].} compared to a collection of local template objects
from the literature (Graci\'a-Carpio et al., in prep.). The FIR luminosity (L$_{FIR}$[40-500$\mu$m]) has been calculated integrating SED data from
the literature (Ao et al. 2008 and references therein) and our PACS continuum measurements, with a lens magnification correction factor of 12. The
upper limit on [O\,IV] is surprisingly low given that the source is classified as an AGN in the optical and mid-IR.
%The [O\,IV] and FIR have both been corrected for the same lens magnification factor of 12.
Teplitz et al. (2006) report a detection of the [Ne\,VI]7.6$\mu$m line in their {\it Spitzer}-IRS spectrum. Assuming an average [Ne\,VI]/[O\,IV]
ratio of 0.4-0.6 in Seyfert galaxies (Sturm et al. 2002) and AGN-ULIRGs (Veilleux et al. 2009) we would have expected a two to three times higher
line flux in [O\,IV] than our upper limit. Only a few objects have a [Ne\,VI]/[O\,IV] ratio larger than 1. Given the uncertainties in both the {\it
Spitzer} and {\it Herschel} measurements our non-detection might still be consistent with a [Ne\,VI] detection. Figure \ref{Fig2}, in any case,
shows that the upper limit on the [O\,IV]/FIR ratio in F10214 is lower than in Seyfert galaxies, and more in the regime of ULIRGs and starbursts.
If the differential magnification model (see section \ref{sect:targets}) is applied, then the upper limit on [O\,IV] is yet lower by a factor of
$\sim$3. The low [O\,IV]/FIR ratio in F10214 is consistent with L$_{FIR}$  arising predominantly from the starburst rather than the AGN, an
assumption that is supported by model fits to the FIR SED, as F10214 has strong emission from the FIR to the (sub-)mm.

\begin{figure}
   \centering
   \includegraphics[width=7cm]{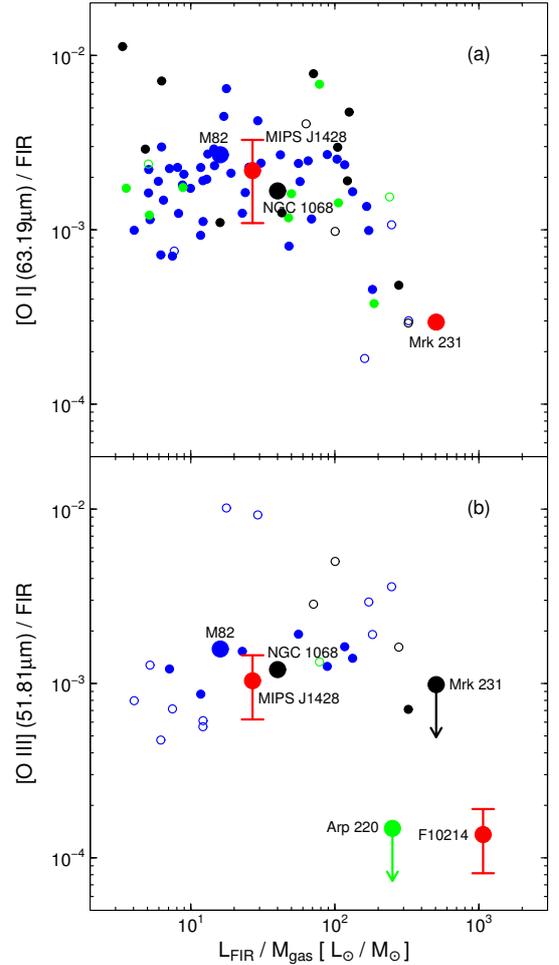}
   \caption{[O I]63$\mu$m and [O III]52$\mu$m line strength in F10214 and MIPS J14284 as function of star formation efficiency.
   Comparison data are from the literature (Graci\'a-Carpio et al., in prep.), except for [O\,I] in Mrk\,231 (Fischer et al. 2010).
   Symbols as in Figure \ref{Fig2}.}
              \label{Fig3}%
    \end{figure}

The [O\,III]/FIR ratio for F10214 is shown in Figure \ref{Fig3}b. It is at least an order of magnitude lower than local starbursts or AGN. There
are not many ULIRG data points available for comparison. We have extracted two upper limits on \object{Arp\,220} and \object{Mrk\,231}, which are
shown in the figure, from ISO-LWS data. Initial results from SHINING (Fischer et al. 2010) indicate a general deficiency of FIR fine-structure
lines (including [O\,III]) in local ULIRGs, similar to the [O\,III] deficiency seen in F10214. With future data from SHINING we will be able to
populate this diagram with many more data points for local templates covering a large parameter space, in order to put the findings reported here
into context.

%______________________________________________ Gamma_1 (lg rho, lg e)

%______________________________________________ Gamma_1 (lg rho, lg e)

{\bf MIPS J1428}: the [O\,I] and [O\,III] lines (Figure \ref{Fig1}) are detected with $\sim$5 $\sigma$ significance. They are redshifted by
220$\pm$100 km/s and 300$\pm$135 km/s, respectively, with respect to z=1.325 (derived from H$\alpha$ and CO). Both lines are resolved with a FWHM
of 750$\pm150$km/s each (corrected for the instrumental profile). This is considerably broader than CO (380$\pm$100 km/s, Iono et al. 2006), but
comparable to H$\alpha$ (530$\pm$160km/s, Borys et al. 2006) within the errors. The measured continuum flux densities at 120 and 150 $\mu$m (Tab.
\ref{table:1}) are low compared to the MIPS 160 $\mu$m flux (430$\pm$90mJy, Borys et al. 2006). However, preliminary processing of PACS and SPIRE
photometry of the source (S. Oliver and H. Aussel, private communication) is consistent with our spectroscopic values. Using the PACS spectrum we
derive L$_{FIR}$(40-500$\mu$m)$\sim$1.3$\times$10$^{13}$L$_\odot$/$\mu$, where $\mu$ is the magnification factor (i.e.
L$_{FIR}\ge$1.6$\times$10$^{12}$L$_\odot$ for $\mu\le$8). Contrary to F10214 the [O\,III]/FIR ratio in MIPS J1428 (Figure \ref{Fig3}) is of the
same order as in local star forming and AGN galaxies. The [O\,III]/[O\,I] ratio is very similar to the ratio in typical starburst galaxies (like
M\,82).

For the [O\,I] spectrum we cannot rule out an underlying absorption of ammonia (NH$_3$). In Figure \ref{Fig1}, top panel, the dotted line is a
NH$_3$ model spectrum (arbitrarily scaled). The line at the bottom of panel (a) is the residual after subtracting the line (+continuum) fit and the
NH$_3$ model from the data. Strong FIR NH$_3$ features with high column densities have been detected before in infrared bright galaxies, in the
infrared (e.g. in Arp\,220, Gonz\'alez-Alfonso et al. 2004) and at mm wavelengths (e.g. Henkel et al. 2008). On the other hand, the corresponding
NH$_3$ column density in MIPS J1428 would be quite high, and the object is not known to be heavily obscured at other wavelengths.

[O\,I] and [C\,II] are usually the brightest cooling lines of the cool ISM in galaxies. However, ISO data have shown the [C\,II] line in local
ULIRGs to be about an order of magnitude lower relative to the FIR continuum than in normal and starburst galaxies (e.g. Malhotra et al. 2001,
Luhman et al. 2003). The relative weakness of [C\,II] has strong implications on its potential use as a star formation tracer for high redshift
(z$\ge$4) studies (e.g. Maiolino et al. 2005 and 2009). Hailey-Dunsheath et al. (2010) have observed [C II] with the ZEUS spectrometer at the CSO.
They found MIPS J1428 to be a counter example, with a normal [C\,II]/FIR ratio. Our [O\,I] observation (Fig. \ref{Fig3}) supports this, i.e. this
source, a high redshift source with the high SFR and L$_{IR}$ of a local ULIRG, does not show a deficit in the major PDR cooling lines relative to
the FIR.

\begin{figure}
   \centering
   \includegraphics[width=9.3cm]{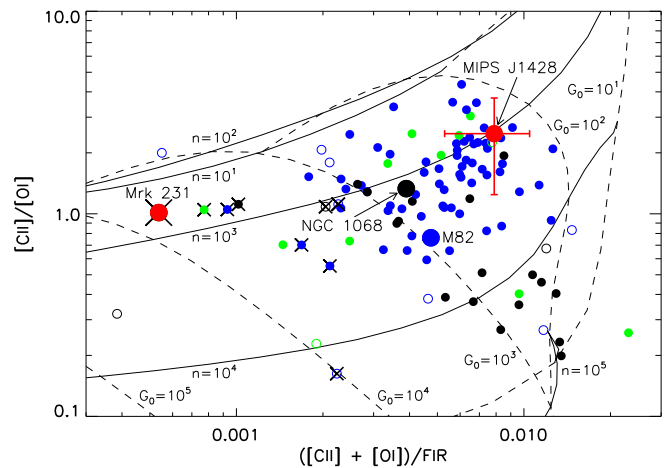}
   \caption{
   [C\,II]/[O\,I] vs. ([C\,II]+[O\,I])/FIR.
   PDR models of Kaufman et al. (1999) are used. The position of MIPS J1428 is indicated, using [C II]
   from Hailey-Dunsheath et al. (2010). Symbols as in Figure \ref{Fig2}, sources with L$_{IR}\ge$10$^{12}$L$_\odot$ are marked with a cross.
   Open symbols correspond to [O\,I] non-detections. The Mrk 231 data are
   from Fischer et al. (2010).}
              \label{Fig5}%
    \end{figure}

We plot in Figure \ref{Fig5} the position of MIPS J1428 in a [C\,II]/[O\,I] versus ([C\,II]+[O\,I])/FIR diagram. For reference, we overplot the PDR
model curves from Kaufman et al. (1999), and show the location of various samples of template galaxies from the literature. This is the first time
that a combined [C\,II] and [O\,I] PDR diagnostic is possible for a source at redshift $\sim$1. The PDR model curves are appropriate for clouds in
the active regions of galaxies that are illuminated on all sides (see Contursi et al. 2002). Combined HII region and PDR modeling of normal and
starbursting galaxies have demonstrated that PDRs account for more than half of the [C\,II] emission in these sources (e.g. Colbert et al. 1999).
Strong shocks could boost the [O\,I] emission relative to [C\,II]. However, the measured [C\,II]/[O\,I] ratio of 2.5$\pm$1.2 is well above 1,
suggesting we can ignore the role of shocks in this case. We conclude that local ULIRGs and AGNs are not good analogs of MIPS J1428, and that this
source appears to be more like a normal/starburst galaxy.

Typical massive galaxies in the distant Universe formed stars an order of magnitude more rapidly than in the local Universe. Either star formation
was significantly more efficient or these young galaxies were much more gas rich (Tacconi et al. 2010). Figure \ref{Fig3} shows the [O\,III]/FIR
ratio as a function of star formation efficiency (SFE, expressed in L$_{FIR}$/M$_{gas}$). M$_{gas}$ was calculated from CO measurements (Iono et
al. 2006, Ao et al. 2008). For F10214 this assumes a CO conversion factor like in local ULIRGs, and  that the FIR arises from star formation rather
than an AGN, an assumption that is supported by the low [O\,IV]/FIR ratio and model fits to the FIR SED (see above). For MIPS J1428 we used a CO
conversion factor for normal star forming galaxies, motivated by our findings above, which yields M$_{gas}$ = 5$\times$10$^{11}$M$_\odot$/$\mu$.
The star formation rate in MIPS J1428 of 2600M$_\odot$/yr/$\mu$ (i.e. SFR $\gtrsim$ 300M$_\odot$/yr for $\mu\le$ 8), is quite high, comparable to
local ULIRGs or to SMGs. However, given the larger gas reservoir, the star formation {\it efficiency} is well in the range of normal star forming
galaxies. Even with a ULIRG-like conversion factor, L$_{FIR}$/M$_{gas}$ would still be lower than in most local ULIRGs. Irrespective of the
conversion factor we find L$_{FIR}$/L$_{CO}$ = 115L$_\odot$/(K km/s pc$^2$) for MIPS J1428. In a recent analysis of CO data from z$\sim$1-3 normal
star forming galaxies Genzel et al. (2010) find values of 5 to 100 for this ratio, while the average ratio for major mergers is $\sim$160, reaching
values up to 600. This is consistent with our interpretation of MIPS J1428 as a normal star forming galaxy. Our result thus adds further support to
the recent results from the SINS survey (Tacconi et al. 2010) which indicate that with an increased gas reservoir star forming galaxies at high
redshifts can achieve ULIRG luminosities without being major mergers.

Upcoming SHINING and other {\it Herschel} observations will enable us to compare with more template objects covering a larger parameter space and
to search for galaxy evolution effects in the ISM of infrared bright galaxies.

%                                     Two column figure (place early!)

%______________________________________________ Gamma_1 (lg rho, lg e)

%______________________________________________________________

\begin{acknowledgements}
PACS has been developed by a consortium of institutes led by MPE (Germany) and including UVIE (Austria); KU Leuven, CSL, IMEC (Belgium); CEA, LAM
(France); MPIA (Germany); INAF-IFSI/OAA/OAP/OAT, LENS, SISSA (Italy); IAC (Spain). This development has been supported by the funding agencies
BMVIT (Austria), ESA-PRODEX (Belgium), CEA/CNES (France), DLR (Germany), ASI/INAF (Italy), and CICYT/MCYT (Spain).
\end{acknowledgements}

%\listofobjects

\end{document}